\documentclass[12pt]{iopart}
\usepackage[superscript]{cite}

\usepackage{setspace}
\usepackage{iopams}  
\usepackage{graphicx}
\usepackage{titlesec}
\usepackage{dcolumn}
\usepackage{amsmath,amssymb}
\usepackage{appendix}
\usepackage{xcolor}
\usepackage{caption}
\usepackage{geometry}
\usepackage{fancyhdr}
\begin{document}
\title{Rotating spokes, potential hump and modulated ionization in radio frequency magnetron discharges}

\author{Liang Xu,$^{1}$ Haoming Sun,$^2$ Denis Eremin,$^3$ Sathya Ganta,$^4$ 
Igor Kaganovich,$^5$
Kallol Bera,$^4$ and Shahid Rauf $^4$} 

\address{$^{1)}$Department of Physical Science and Technology, Soochow University, Suzhou 215006, China}
\address{$^{2)}$École polytechnique fédérale de Lausanne (EPFL), Swiss Plasma Center (SPC), CH-1015 Lausanne, Switzerland}
\address{$^{3)}$Institute for Theoretical Electrical Engineering, Ruhr University Bochum, D-44801 Bochum, Germany}
\address{$^{4)}$Applied Materials, Inc., USA}
\address{$^{5)}$Princeton Plasma Physics Laboratory, USA}

\vspace{5pt}
\begin{abstract}
In this work, the rotating spoke mode in the radio frequency (RF) magnetron discharge, which features the potential hump and the RF-modulated ionization, is observed and analyzed by means of the two dimensional axial-azimuthal ($z-y$) particle-in-cell/Monte Carlo collision method. The kinetic model combined with the linear analysis of the perturbation reveals that the cathode sheath (axial) electric field $E_{z}$ triggers the gradient drift instability (GDI), deforming the local potential until the instability condition is not fulfilled and the fluctuation growth stops in which moment the instability becomes saturated. The potential deformation consequently leads to the formation of the potential hump, surrounding which the azimuthal electric field $E_{y}$ is present. The saturation level of $E_{y}$ is found to be synchronized with and proportional to the time-changing voltage applied at the cathode, resulting in the RF-modulation of the electron heating in the $E_{y}$ due to $\nabla B$ drift. In the instability saturated stage, it is shown that the rotation velocity and direction of the spoke present in the simulations agree well with the experimental observation. In the instability linear stage, the instability mode wavelength and the growth rate are also found to be in good agreement with the prediction of the GDI linear fluid theory.

\end{abstract}


%
%
%
%

\ioptwocol
\section{Introduction}
Azimuthal spoke modes ubiquitously occur in magnetron plasmas with all discharge types: direct current (DC), pulse and radio frequency (RF) driven magnetrons \cite{panjan2015non,kozyrev2011optical,ehiasarian2012high,anders2012drifting,poolcharuansin2015use,hnilica2018effect,brenning2013spokes,panjan2019self}, as well as in other typical $\bf E\times \bf B$ plasmas, such as Hall thrusters \cite{janes1966anomalous,MazouffrePSST2019,sekerak2014azimuthal,griswold2012feedback,esipchuk1974plasma} and Penning discharges \cite{thomassen1966turbulent,sakawa1993excitation,rodriguez2019boundary,gonzalez2020spatially}. The spoke mode significantly affects the dynamics of electrons and ions and its fundamental understanding is hereby of importance in the operation and control of these $\bf E\times \bf B$ plasmas based applications \cite{sengupta2021restructuring,Boeuf2020,koshkarov2019self,kawashima2018numerical,Xu2021,sakawa1993excitation,smolyakov2016,fridman1964phenomena,lakhin2018effects,lucken2019instability,hara2022theory}. In magnetrons, the spoke refers to a locally enhanced ionization zone rotating in the azimuthal ($\bf E\times \bf B$) direction above the erosion area (i.e., race track) where the target (cathode) surface is sputtered the most. One prominent feature of spokes experimentally observed in magnetron discharges is the high potential region compared to its surrounding regions, so-called potential hump \cite{panjan2014asymmetric,panjan2017plasma}. This local potential structure was assumed to play an important role on the electron heating responsible for the enhanced ionization defining the spoke. Recently, the electron heating due to the local $\nabla B$ drift at the leading edge of the spoke (or the potential hump) was proposed by Boeuf \cite{Boeuf2020}, having addressed the locally enhanced ionization of the spoke in a DC magnetron plasma. The $\nabla B$ drift-induced energy gain of electrons in the electric field is written as

\begin{align}
    \frac{\partial \varepsilon_{\perp}}{\partial t}=-{\bf v_{\nabla B}} \cdot {\bf E}= \frac{\varepsilon_{\perp}E_{y}}{B|L_B|}
\end{align}

\noindent where $v_{\nabla B}=\varepsilon_{\perp}/BL_B$ is the $\nabla B$ drift velocity, $L_B=B/\frac{dB}{dz}$ is the magnetic field gradient length in the axial ($z$) direction, $\varepsilon_{\perp}$ is the electron kinetic energy perpendicular to the magnetic field, $E_{y}$ is the azimuthal electric field due to the presence of the spoke. $E_{y}$ can be either positive or negative corresponding to the electron heating (associated with the spoke ionization) or cooling respectively. Although spokes in DC and pulsed magnetrons were studied extensively, but their presence and the underlying physics are little studied in RF magnetron discharge, which is an important technique for the deposition of both insulating and conducting films for semiconductor manufacturing \cite{Carcia2003,Yabuta2006,Fortunato2010,Lee2020}. The first experimental evidence of the rotating spoke in RF magnetron discharges was recently reported by Panjan\cite{panjan2019self}, but its formation and physics is not elucidated by far. In this work, the locally enhanced ionization of spoke in RF magnetron discharge, which is found to be RF-modulated, is identified and explained by means of 2D PIC/MCC method.

One mystery of the spoke is the formation mechanism of the spoke potential structure-potential hump. Previous numerical works suggested that a spoke mode is originated from the drift instability driven by the gradients of potential and density \cite{Simon1963,hoh1963,sakawa1993excitation}. Various theoretical modifications of the gradient drift instability (GDI) were developed, taking into account a variety of effects, such as the magnetic field gradient, inertia, collisions, ion beam and ionization \cite{smolyakov2016,lakhin2018effects,Ito2015,hara2022theory,lucken2019instability}. The evidence of GDI being the origin of the spoke mode was identified by particle-in-cell (PIC) simulations in planar magnetrons \cite{Boeuf2020}, cylindrical magnetrons \cite{sengupta2021restructuring, Xu2021} and Penning-type discharges \cite{powis2018scaling}. Fluid modeling \cite{koshkarov2019self} and hybrid simulation \cite{kawashima2018numerical} in the configuration of Penning-type discharges also showed that GDI can evolve into the spoke mode. Some of the numerical simulations clearly presented that in the saturated nonlinear stage, the large scale spoke with a locally enhanced ionization region forms a potential hump, which is consistent with the numerous experimental observations in magnetron discharges. However, the question on how the linear perturbations of GDI transit to the formation of the spoke potential hump is still not clear. This question is another focus of the present paper.

We carried out simulations of the recently published experiment by Panjan \cite{panjan2019self} related to an RF magnetron discharge at low pressures $\rm P=0.5-2 \,{\rm Pa}$, where rotating spokes with mode number $\rm m=1-5$ were observed. The experiment consists of a planar RF magnetron discharge in argon (electrode gap size $2 \,{\rm cm}$, driven frequency $13.56 \,{\rm MHz}$, magnetic field $70 \,\rm mT$ at the cathode) where the light emission is observed by a fast camera (ICCD). In the experiment, about seven frames per second were recorded by the camera meaning the time interval of two frames is about million RF periods and thereby the RF period ($T\approx74 \,{\rm ns}$) is not resolved \cite{panjan2023}. However, the resolution of the plasma in the single RF cycle is essential to unravel the electron dynamics behind the spoke formation. Our 2D PIC/MCC simulations under conditions of this experiment enable a resolution of the plasma parameters of one full RF period, and hereby capture and characterize the response of the spoke and electron dynamics to the applied radio frequency. In this paper, section 2 outlines the numerical model, section 3 details the main results and discussions, and the work is concluded in section 4.

\begin{figure*}
\center
\includegraphics[clip,width=1.0\linewidth]{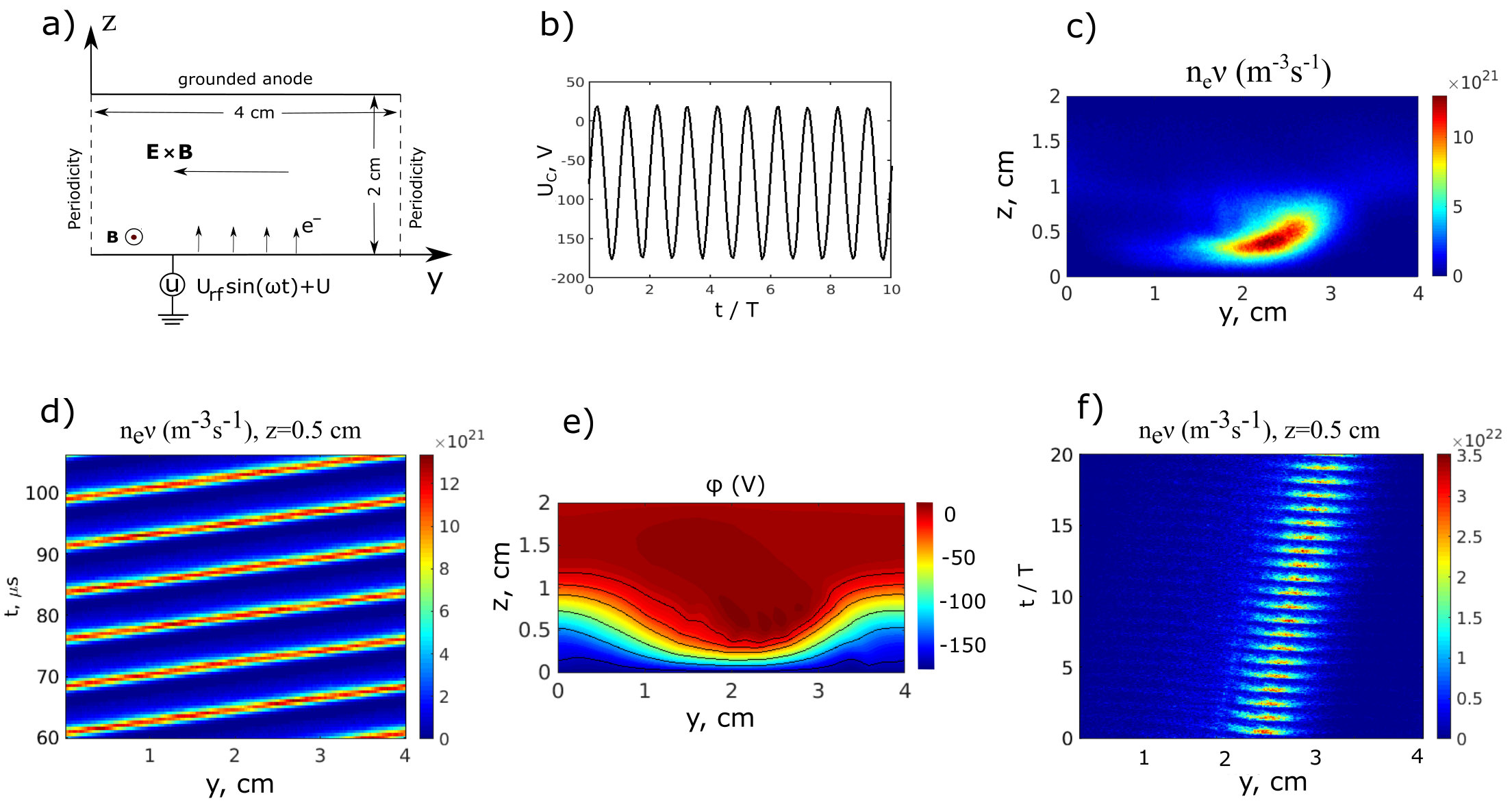}
\caption{a) The layout of the 2D axial ($z$)-azimuthal ($y$) PIC/MCC model of the RF magnetron plasma. b) The voltage waveform applied at the cathode. c) The 2D map of the RF period averaged ionization rate.  d) The spatio (azimuthal)-temporal plot of the ionization rate in the long duration at $z=0.5 \,{\rm cm}$ (note the ionization rate is RF period averaged). Animations are provided in the supplementary materials to
visually demonstrate the spatio-temporal evolution of RF-period averaged ion density and ionization rate. e) The contour 2D map of potential in the steady state when cathode voltage is minimum ($U_c=-180 \,{\rm V}$). f) The spatio (azimuthal)-temporal plot of the ionization rate in the duration of 20 RF periods at $z=0.5 \,{\rm cm}$ (here the ionization rate is transient instead of the RF period-averaged).}
\end{figure*}

\section{Numerical model}
To conduct the simulations, we use 2D-EDIPIC, an electrostatic momentum conserving explicit 2d3v PIC/MCC code, which was benchmarked against various codes \cite{edipic,charoy2019,villafana2021}. We map the cylindrical geometry of magnetron onto a Cartesian coordinate system, and the axial ($z$) and azimuthal ($y$) dimensions are resolved at the radial position where the spoke fluctuation is strong, i.e., at the race track. The working gas is argon and the pressure is $P=2 \,{\rm Pa}$. The external magnetic field ${\bf B_x}$ is taken along the $x$-direction (radially outward) and the electric field ${\bf E_z}$ is oriented along the $z$-direction (positive from cathode to anode) so that electrons drift azimuthally in $y$ ($\pm {\bf E_z}\times {\bf B_x}$) direction. The simulation box area is $L_{y}\times L_{z}=4 \,{\rm cm}\times 2 \,{\rm cm}$. The $L_{y}=4 \,{\rm cm}$ is chosen, because the length of one spoke observed in the experiments when $P=2 \,{\rm Pa}$ is about $4 \,{\rm cm}$. The azimuthal boundaries are periodical. Similar to the work in \cite{Boeuf2020}, the external magnetic field is set to be $B_x(z)=a exp(-z^2/2\sigma^2)+b$ with $\sigma=0.35L_z$, where $a$ and $b$ are chosen so that $B_x(z=0)=70\,{\rm mT}$ and $B_x(z=L_z)=0$. The cathode is driven by the RF voltage with the frequency $\omega_{rf}=13.56\,{\rm MHz}$. The peak to peak voltage amplitude is $200\,{\rm V}$ with a shifted DC voltage $-80 \,{\rm V}$ to mimic the self-bias voltage in the experiment due to the geometrical asymmetry. The simulation box has $256 (y) \times 128 (z)$ cells giving the cell size $\triangle y=\triangle z=0.15 \,{\rm mm}$ and the time step is $50\,{\rm ps}$, so that the plasma frequency and Debye length are resolved. The initial state of the simulation is a stationary Maxwellian distribution for both electrons and ions with a homogeneous plasma background ($n_0=5\times 10^{15}\,{\rm m^{-3}}$). The ions impinging the cathode can knock out secondary electrons with the emission coefficient $\gamma=0.1$. The elastic, excitation and ionization electron-neutral collisions and charge exchange ion-neutral collision are implemented. The electron-neutral collision cross sections are those of Phelps \cite{phelps1999cold} and the charge exchange cross section is set to be $5.53\times 10^{-19} \,{\rm m^{2}}$. The model layout and the time-changing driven voltage are illustrated in Fig. 1a and 1b. The simulations start with the number of particles per cell $N_{ppc}=400$ and reach the steady state at approximately $t=10 \,{\rm \mu s}$ with $N_{ppc}\approx 200$. 

\begin{figure}
\center
\hspace*{0.2cm}\includegraphics[clip,width=0.9\linewidth]{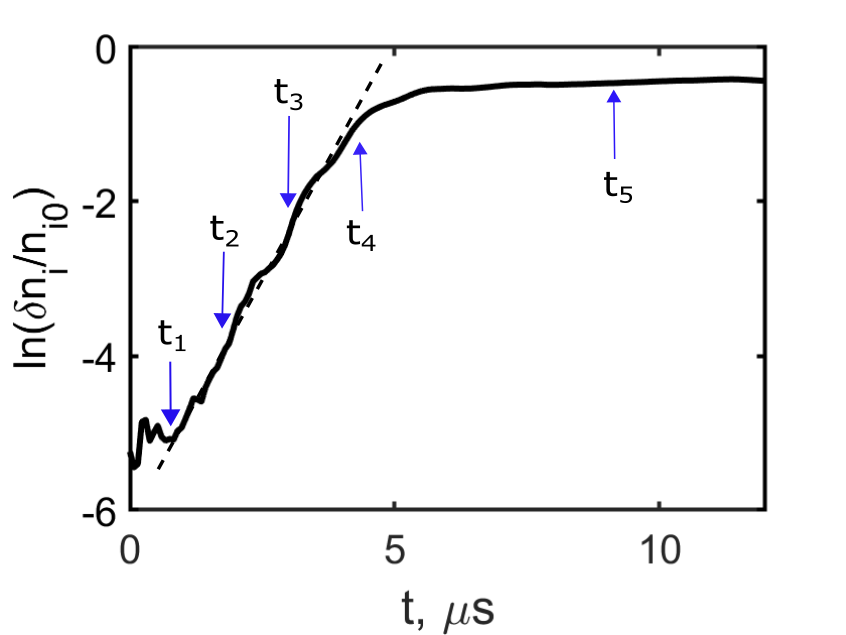}
\caption{The ion density fluctuation as a function of time and $t_1 \rightarrow t_5$ are selected snapshots to trace the instability development.}
\end{figure}

\section{Results and discussions}
In the nonlinear saturated stage, our numerical simulation exhibits a locally enhanced ionization zone as shown in the 2D map of the ionization rate $n_e\nu$ in Fig. 1c. It is noted that the ionization rate is RF period averaged. Fig. 1d gives the temporal-spatial (azimuthal) evolution of $n_e\nu$ at $z=0.5 \,{\rm cm}$ showing that the spoke rotates in the $-\bf E_z\times \bf B_y$ direction with the mode number of $m=1$ and the velocity of $5\,\rm km/s$, consistent with the experimental data \cite{panjan2019self}. An interesting observation is that, if the RF period is resolved, the ionization rate exhibits oscillation with the applied radio frequency as shown in Fig. 1f, which gives the temporal-spatial (azimuthal) evolution of $n_e\nu$ in 20 RF periods. For Fig. 1f, we should emphasize that the $n_e \nu$ is transient (not averaged) data, and the $n_e \nu$ data were sampled at $z=0.5 \,{\rm cm}$. The contour plot of the transient potential in Fig. 1e (when the cathode voltage is minimum $-180 \,{\rm V}$) clearly presents the potential hump region, breaking the azimuthal symmetry, deforming the equi-potential lines and reminding of the potential measurements in DC and pulsed magnetron discharges \cite{panjan2014asymmetric,panjan2017plasma}. In what follows, the underlying physics of occurrences of the spoke potential hump and the RF-modulated spoke ionization will be explained to address the driving mechanism of rotating spoke in the context of RF magnetron discharges. 

\begin{figure*}
\center
\includegraphics[clip,width=1.0\linewidth]{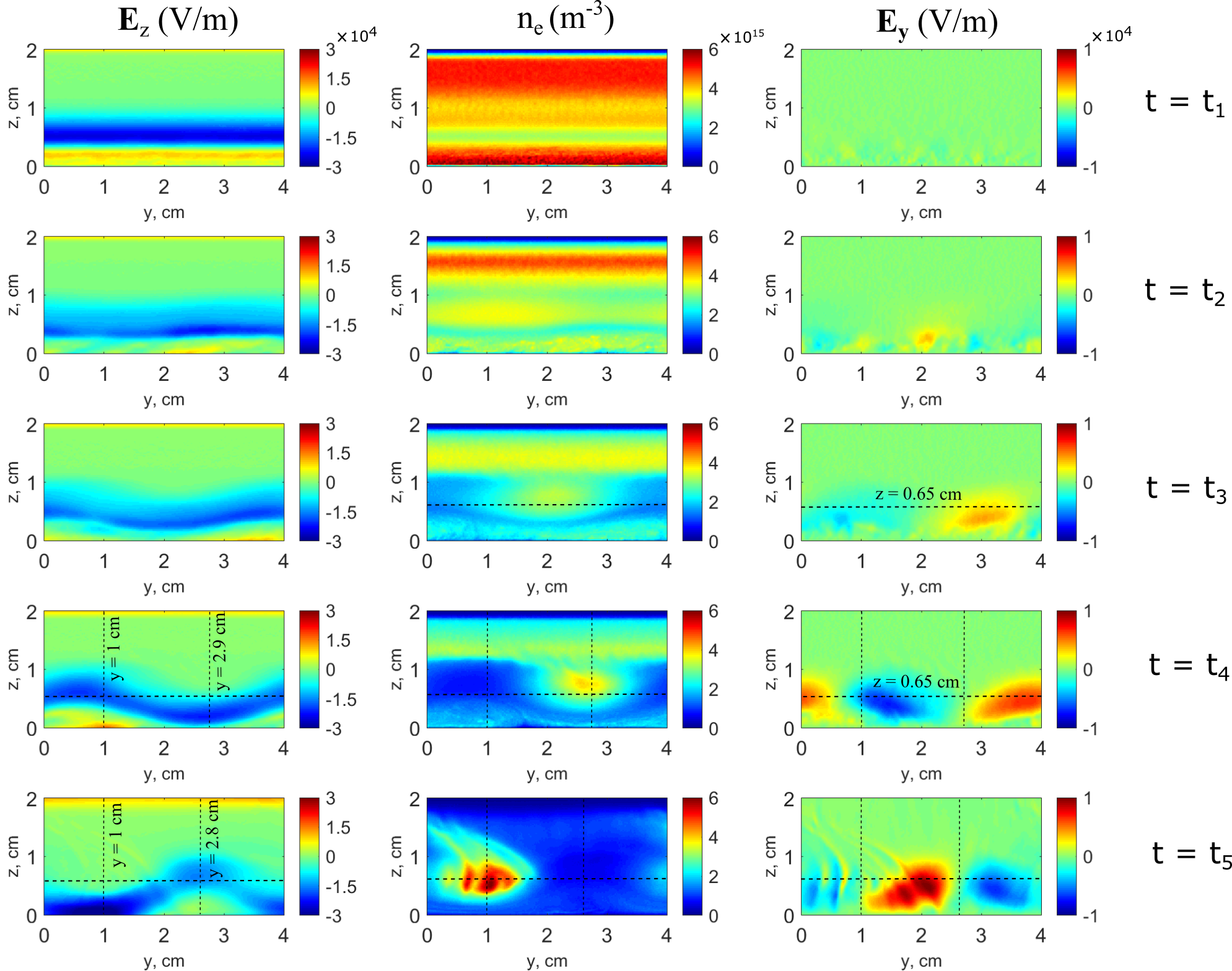}
\caption{Two dimensional maps of $E_z$, $n_e$ and $E_{y}$ at each snapshot. The data along the dashed lines at $t=t_3$, $t=t_4$ and $t=t_5$ are chosen to check the instability criteria locally.}
\end{figure*}

To address the transition from the micro fluctuations to the formation of the spoke potential hump, the fluctuations and the equilibrium parameters in the early phase of the computational simulation are monitored to identify the instability onset and the nonlinear evolution. Shown in Fig. 2 is the history of ion density fluctuation $\delta n_i/n_{i0}$ in the logarithm scale during the time range $[0, 15\mu s]$. To specify, here $\delta n_i=[\int^{L_{z}'}_{0}\int^{L_{y}}_{0} n_i(z,y,t)^2 d z d y-(\int^{L_{z}'}_{0}\int^{L_{y}}_{0}n_i(z,y,t) d z d y)^2]^{1/2}$, and 

$n_{i0}=\int^{L_{z}'}_{0}\int^{L_{y}}_{0} n_i(z,y,t) d z d y$. Here, $L_{z}'=1 \,{\rm cm}$ is used because the spoke fluctuation takes place at $z=[0,1 \,{\rm cm}]$ (see Fig. 3 below). We should emphasize that the RF frequency is much larger than the instability frequency and the growth rate ($\omega_{rf}\gg \omega/2\pi, \gamma/2\pi$), implying that the DC voltage component prevails over the RF oscillation component on the instability development (see the appendix). So the ion density used for Fig. 2 plot is RF period averaged. As clearly shown in Fig. 2, the instability grows linearly from $t\approx 0.5 \,{\rm \mu s}$ to $t \approx 4.5 \,{\rm \mu s}$, after which the fluctuation stops the linear growth, i.e., enters the nonlinear stage. To trace the instability development, five snapshots are selected: $t_1=0.5 \,{\rm \mu s}$ is the moment when the instability is not excited yet; $t_2=2.25 \,{\rm \mu s}$ and $t_3=3.75 \,{\rm \mu s}$ are the moments when the fluctuation grows linearly; from $t_4=4.5 \,{\rm \mu s}$ on, the fluctuation starts to deviate from linear growth and enter the nonlinear stage; at $t_5=9.0 \,{\rm \mu s}$, the growth fully stops and the instability saturates. At each snapshot, the $z-y$ distributions of $E_z$, $n_e$ and $E_y$ are presented in Fig. 3. The instability is presumed to be gradient drift instability. For the analysis of the perturbation present in our simulations, it is convenient to start with the linear dispersion relation of GDI derived via two-fluid theory, which reads \cite{lakhin2018effects, smolyakov2016},

\begin{equation}
 \begin{aligned}
1-\frac{\omega_{pi}^2}{\omega^2}+\frac{\omega_{pe}^2}{\omega_{ce}^2}\frac{1}{1+k_{y}^2\rho_e^2}+\frac{1}{k_{y}^2\lambda_{De}^2} \\
\frac{1}{1+k_{y}^2\rho_e^2}\frac{k_{y}v_{*}-\beta k_{y}v_{\nabla B}}{\omega-\beta k_{y}v_{\nabla B}-k_{y}v_{E}} =0 
\end{aligned}
\end{equation}

\noindent where $k_{y}$ is the angular wave number in the azimuthal direction, $\omega$ the angular frequency, $\lambda_{De}$ the Debye length, $\rho_{e}=(T_e/m_e)^{1/2}/\omega_{ce}$ the electron Larmor radius, $m_e$ the electron mass, $T_e$ the electron temperature, $\omega_{ce}$ the electron cyclotron frequency, $\omega_{pe}$ the electron plasma frequency, $\omega_{pi}$ the ion plasma frequency, $m_i$ the ion mass, diamagnetic drift velocity ${\bf v}_{*}=-T_e/(eB_x)\nabla ln n_{e0}$, $\nabla B_x \times {\bf B}_x$ drift velocity ${\bf v}_{\nabla B}=-2T_e/(eB_x)\nabla ln B_x$, ${\bf E} \times {\bf B}_x$ drift velocity ${\bf v}_E=-{\bf E}_0/B_x$, $\beta$ the coefficient $(1+2k_y^2\rho_e^2)/(1+k_y^2\rho_e^2)$ and $e$ the elementary charge. Here, $E_0$, $B_x$, $n_{e0}$ are the equilibrium electric field, magnetic field and electron density, and $k_{y}$ is in the unit of ${\rm rad/m}$ and $\omega$ in ${\rm rad/s}$. The first term on the left hand side of Eq. 2 refers to the Debye length effect derived from the Poisson equation, the second term is attributed to the ion inertial response and the last two terms on the left hand are related to the electron response including the inertia and the gyroviscosity. We note again that the RF oscillating electric field is negligible regarding the derivation of the GDI dispersion relation and $E_0$ in Eq. 2 can be approximated by the RF period averaged value (see the appendix on addressing the negligible effect of RF oscillation on GDI theory). It's also noted that Eq. 2 is obtained under the condition $k_{y}\gg k_{z}$ and the assumption of local theory $k_{y}|L_n| \gg 1 $, where $k_{z}$ is the wave number in the axial direction, $L_n=n_{e0}/\frac{dn_{e0}}{dz}$ is the electron density gradient length. The condition $k_{y}\gg k_{z}$ is valid concerning the purely azimuthal spoke mode. The assumption of the theory to be local is loosely fulfilled due to the condition of $k_{y}|L_n| \gtrsim 1 $ in our simulations, which may bring nonlocal effect and hereby discrepancy of the comparison between the simulation and the local theory.


\begin{figure}
\center
\includegraphics[clip,width=0.7\linewidth]{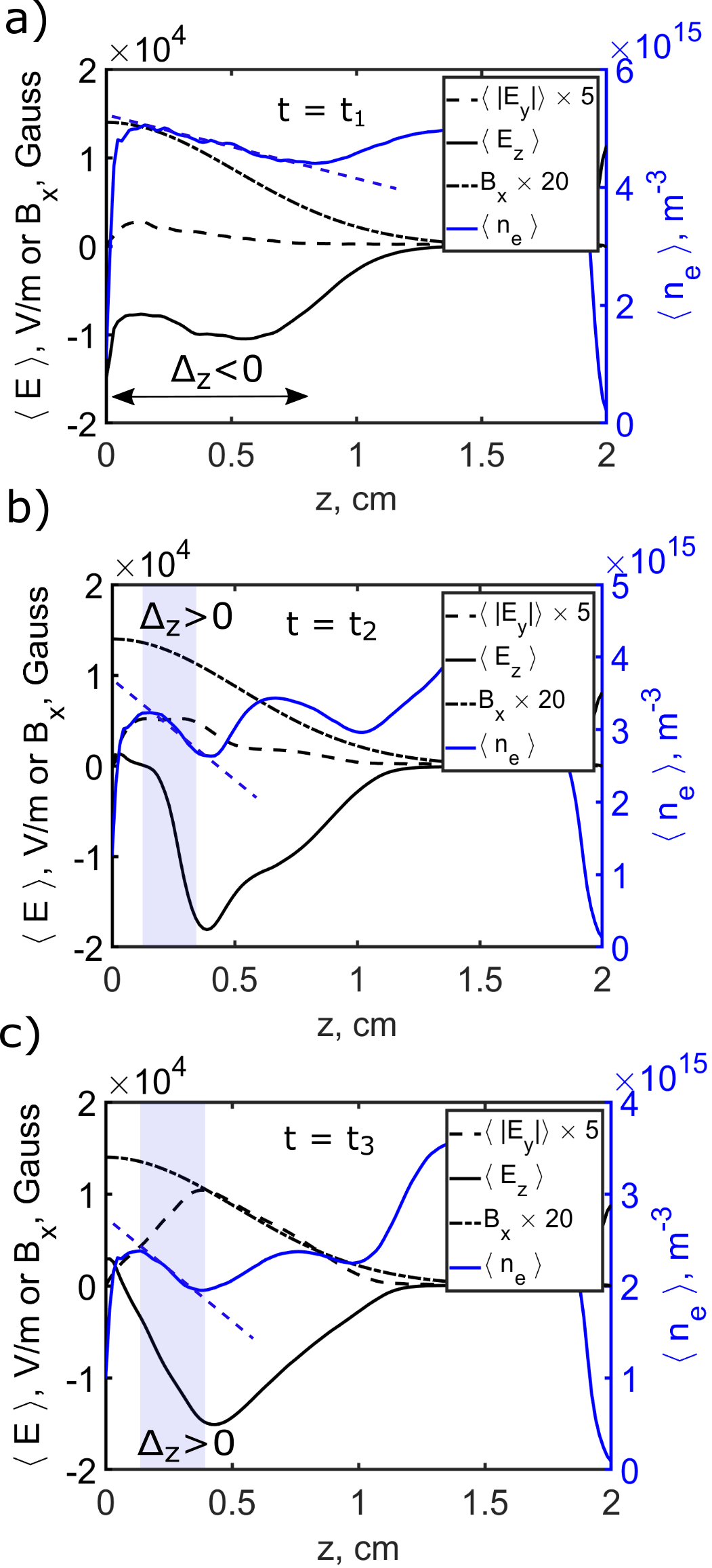}
\caption{The axial profiles of azimuthally averaged $E_z$, $n_e$, $E_{y}$ and $B_x$ a) at $t=t_1$ (in the prelinear stage), b) at $t=t_2$ and c) at $t=t_3$ (both in the linear stage).}
\end{figure} 

For the linear stage concerning the azimuthal uniformity,  the three velocities in Eq. 2 become $v_{*}=-T_e/(eB_xL_n)$, ${v_E}=-{E_{0z}}/B_x$ and $v_{\nabla B}=-2T_e/(eB_xL_B)$, meaning the axial gradients of plasma density, potential and magnetic field are critical to determine the linear characteristics of the instability. Here, $E_{0z}$ is the $z$ component of the equilibrium electric field $E_0$. To calculate the dispersion relation using Eq. 2, the critical parameters $E_{0z}$, $L_B$, $L_n$ are therefore required as input and can be estimated from the PIC simulations. In the linear (at $t=t_2$ and $t=t_3$) and pre-linear (at $t=t_1$) stages, Fig. 4 shows the axial profiles of azimuthally averaged equilibrium parameters of $\langle E_{z} \rangle$ and $\langle n_e \rangle$ to approximate $E_{0z}$ and $L_n$. Meanwhile,  Fig. 4 also presents $\langle |E_{y}|\rangle$ to identify the location where fluctuation arises and $B_x$ to get $L_B$.  As shown in Fig. 4b and 4c, the parameters in the shaded regions where $\langle |E_{y}|\rangle$ are roughly peaked (namely instability occurs) are of interest. In the shaded region of Fig. 4b and 4c, $L_n \approx -0.7 \,{\rm cm}$, $L_B \approx -1.6 \,{\rm cm}$ can be derived. But $\langle E_z \rangle$ is strongly spatially nonuniform, so two values of (lower limit) $E_{0z}=\langle E_z \rangle=-2 \,{\rm kV/m}$ and (upper limit) $E_{0z}=\langle E_z \rangle=-15 \,{\rm kV/m}$ in the shaded region are taken from Fig. 4b and 4c to calculate the dispersion relation. Inserting the parameters to Eq. 2, the calculated theoretical predictions of the real part and the imaginary part (growth rate) of the frequency as a function of the mode number $k_{y}L_{y}/2\pi$ is displayed in Fig. 5. From the growth rate plots of Fig. 5 when $E_{0z}=-2 \,{\rm kV/m}$, the maximum growth rate $\gamma_{max}=1.0\times 10^7 \,{\rm rad/s}$ is calculated and its corresponding mode (the most unstable one) has mode number $m\approx 1$. On the other hand, the simulated growth rate $\gamma_{pic}=1.05\times 10^7 \,{\rm rad/s}$ is obtained from Fig. 2 by fitting the linear growth. So the theoretical prediction of the mode number and the growth rate agrees very well with the simulation when $E_{0z}=-2 \,{\rm kV/m}$. Likewise, from Fig. 5 when $E_{0z}=-15 \,{\rm kV/m}$, $m\approx 0.4$ and $\gamma_{max}=1.5\times 10^7 \,{\rm rad/s}$ are identified, two of which also do not differ much from the observations in the simulation. The reason for the discrepancy between theory and simulation could be twofold. First, the equilibrium parameters of $\langle E_z \rangle$ and $\langle n_e \rangle$ are both temporally and spatially dependent in the linear phase. Second, in our case, the condition of $k_{y}|L_n| \gtrsim 1$ leads to the assumption of the local theory of Eq. 2, $k_{y}|L_n| \gg 1$, not strictly valid and hereby the possible introduction of the nonlocal effect. In our case, the spoke rotation frequency $f_{pic}=7.5 \times 10^5 \,{\rm rad/s}$ is obtained in the nonlinear stage, which is lower than the theoretical real frequency $\omega=1\times 10^6 \,{\rm rad/s}$ for $E_{0z}=-2 \,{\rm kV/m}$ case and $\omega=2\times 10^6 \,{\rm rad/s}$ for $E_{0z}=-15 \,{\rm kV/m}$ case (the real frequency corresponds to the most unstable mode). This discrepancy is attributed to the limit of linear theory in predicting nonlinear behavior. The spoke rotation frequency is marked as the squared points in Fig. 5.

\begin{figure}
\center
\includegraphics[clip,width=0.9\linewidth]{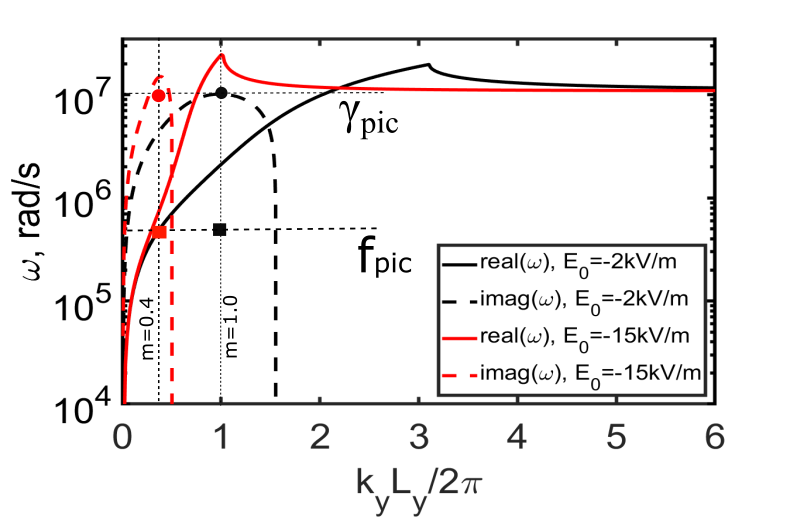}
\caption{The calculated dispersion relation by Eq. 2 using the equilibrium parameters obtained from Fig. 4b and 4c, $E_0=-2\times 10^3 \,{\rm V/m}$ or $E_0=-1.5\times 10^4 \,{\rm V/m}$, $L_n=0.7 \,{\rm cm}$ and $L_B=1.6 \,{\rm cm}$. The upper horizontal line denotes $\omega=\gamma_{pic}$ which is the growth rate derived from Fig. 2. The bottom one is $\omega=f_{pic}$, that is the spoke rotation frequency in the nonlinear saturated stage. The vertical dashed lines represent the most unstable modes having the largest growth rates.}  
\end{figure}

Besides the dispersion relation, the linear analysis can also be conducted by checking the instability criteria of GDI \cite{lakhin2018effects}, which writes,

\begin{equation}
 \begin{aligned}
\triangle=\triangle_z+\triangle_{y}>0\\
 \end{aligned}
\end{equation}

\begin{equation}
   \triangle_z =(\frac{eE_{0z}}{T_e}+2/L_B)(1/L_n-2/L_B)
\end{equation}

\begin{equation}
   \triangle_{y}=\frac{eE_{0y}}{T_e} \frac{1}{n_{e0}}\frac{d n_{e0}}{d y}
\end{equation}

\noindent where $E_{0z}$ and $E_{0y}$ are the axial and azimuthal components of the equilibrium electric field. We note $\triangle_{y}$ does not include the contribution of the magnetic field due to the azimuthal uniformity of $B_x$. If the magnetic field is further assumed to be axially uniform, Eq. 3 turns out to be $\bf E \cdot \nabla n_{e0}>0$, which is the well-known instability criteria of the long wavelength gradient drift instability, i.e., Simon-Hoh instability \cite{Simon1963,hoh1963,sakawa1993excitation}. As pointed out above, in the linear stage, azimuthal uniformity of $n_e$ can be assumed, hence $\triangle_{y}=0$ and the parameters in Eq. 4 ($L_n$, $L_B$ and $E_{0z}$) are again approximated via azimuthally averaged values from the shaded regions in Fig. 4b and 4c. Then $E_{0z}<0$, $L_n\approx -0.7 \,{\rm cm}$ and $L_B \approx -1.6 \,{\rm cm}$ are obtained, meaning $1/L_n-2/L_B<0$ and $\triangle >0$ at $t=t_2$ and $t=t_3$. Therefore, the instability is triggered and the fluctuation grows at these two moments. Correspondingly, the fluctuations of $n_e$ and $E_{y}$ shown in their 2D maps of Fig. 3 are visible at $t=t_2$ and $t=t_3$. Whereas at $t=t_1$ in the near-cathode region of Fig. 4a where $L_n\approx 2.5 \,{\rm cm}$ and $1/L_n-2/L_B<0$, $E_{0z}>0$ are seen, meaning $\triangle<0$ and the instability is not triggered. We further noticed that at $t=t_3$, the azimuthal non-uniformity of $n_e$ and $E_{y}$ in Fig. 3 becomes pronounced, which is likely to result in non-zero $\triangle_{y}$ and hence cause local instability. Fig. 6 gives the azimuthal profiles of $E_{y}$ and $n_e$ at $t=t_3$ and at $z=0.65 \,{\rm cm}$ where the fluctuation amplitude, i.e., $|E_{y}|$, is largest. We should emphasize that along this azimuthal line, $L_n>0$ and $E_{0z}>0$ (see Fig. 5c at $z=0.65 \,{\rm cm}$) and thereby $\triangle_z<0$. Therefore, the local stability condition can be checked by the sign of $\triangle_{y}$. In Fig. 6, $M$ denotes the cross point of the vertical line of $n_e$ peak and the horizontal line of $E_{y}=0$; $N$ denotes the cross point of the vertical line of $n_e$ trough and the horizontal line of $E_{y}=0$. It is very interesting to see that $M$ and $N$ fall onto the $E_{y}$ curve. One can easily tell that on the left side of $M$, $\frac{d n_e}{d y}>0$ and $E_{y}<0$, i.e., $\triangle_{y}<0$; on the right side of $M$, $\frac{d n_e}{d y}<0$ and $E_{y}>0$, i.e., $\triangle_{y}<0$. It is also seen that $\triangle<0$ holds for the two sides of $N$. Therefore, at $t=t_3$, the azimuthal non-uniformity caused by the original GDI gives negative $\triangle_{y}$ and does not lead to the local "secondary" GDI in the linear stage. Another important observation is that in the 2D maps of $E_{z}$ in Fig. 3, there forms a channel where $E_{z}$ is negatively large and thereby electrons drift in the azimuthal $\bf E_z \times \bf B_x$ channel. In the linear stage, the channel keeps almost straight azimuthally, meaning the potential deformation is not strong.

\begin{figure}
\hspace{0cm}\includegraphics[clip,width=0.8\linewidth]{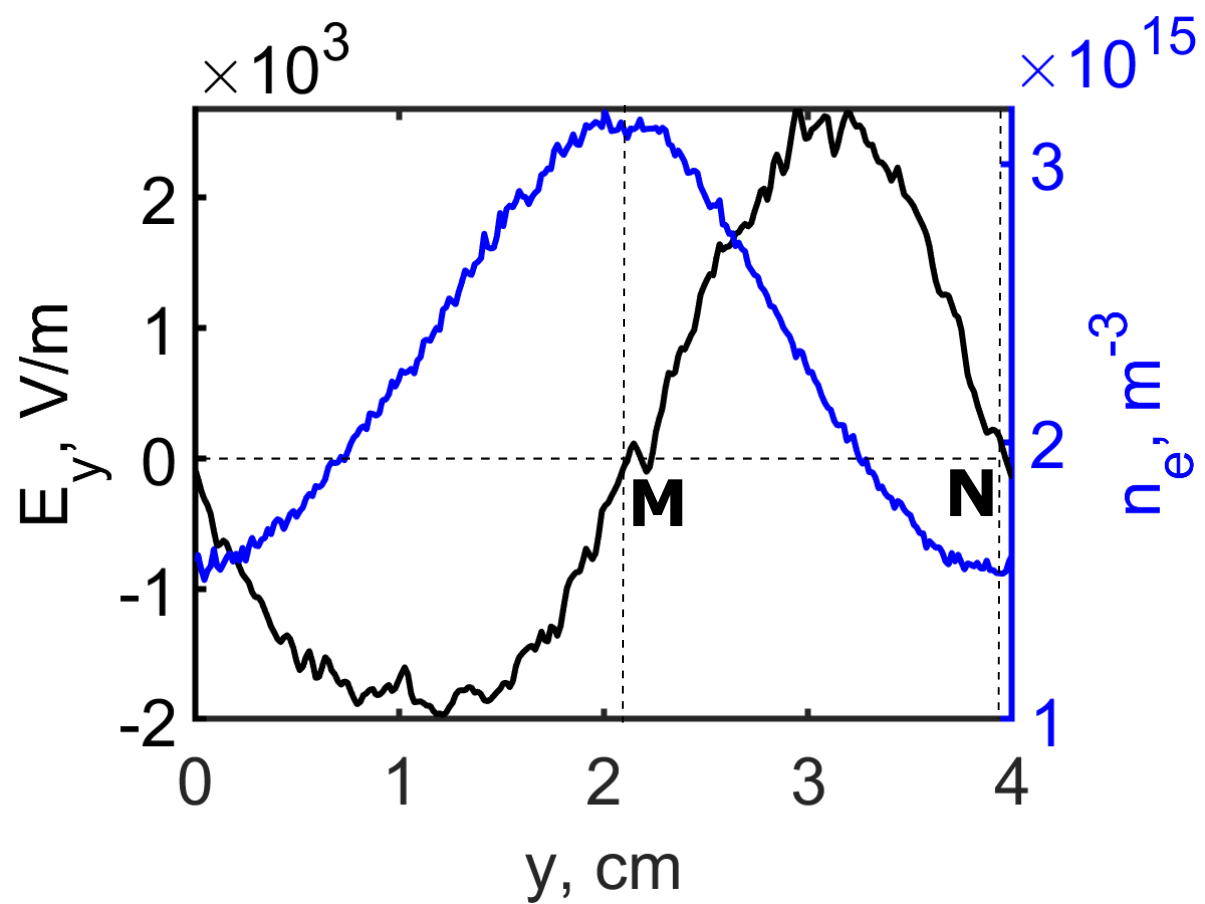}
\caption{The azimuthal profile of $E_{y}$, $n_e$ at $t=t_3$ and at $z=0.65 \,{\rm cm}$.}
\end{figure}

\begin{figure}
\includegraphics[clip,width=1.0\linewidth]{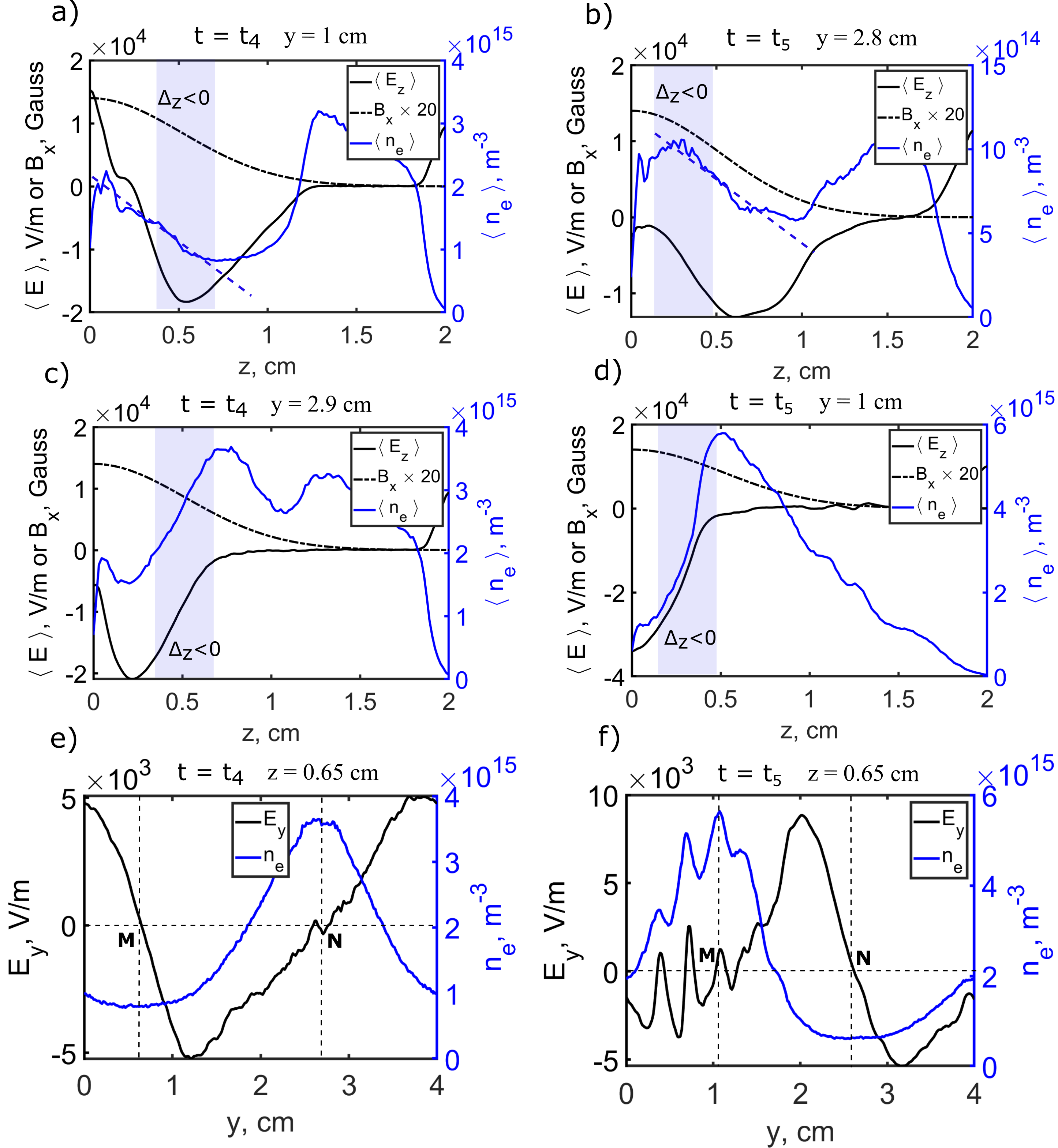}
\caption{a), b), c) and d) show the axial profiles of $E_z$, $n_e$, $E_{y}$ and $B_x$ at the azimuthal locations of $n_e$ peak and $n_e$ trough, and at the two snapshots $t=t_4$ and $t=t_5$. e) and f) are the azimuthal profiles of $E_{y}$ and $n_e$ at $z=0.65 \,{\rm cm}$ and at the two snapshots $t=t_4$ and $t=t_5$.}
\end{figure} 

\begin{figure*}
\center
\includegraphics[clip,width=1.0\linewidth]{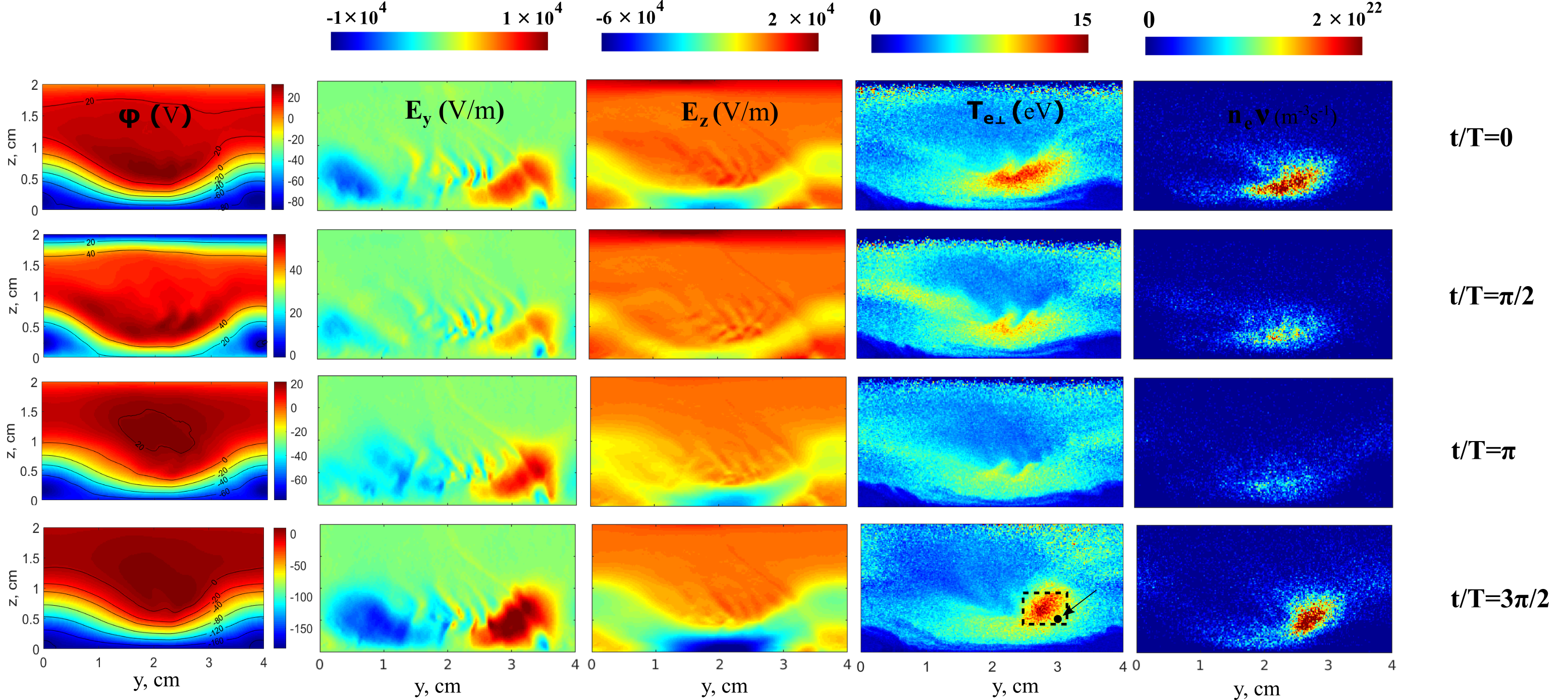}
\caption{The resolution of $z-y$ distributions of $\varphi$, $E_{y}$, $E_z$, $T_{e\perp}$ and $n_e\nu$ in one RF period in the steady state. The time moments can be seen in Fig. 9. The small box is the region, over which the electrons representing EVDFs in Fig. 9b are collected. The point in the box is chosen to show the temporal evolution of $E_{y}$ and $|E_{z}|$ in Fig. 9a. }
\end{figure*} 
 
At $t=t_4$ and $t=t_5$ when the instability enters the nonlinear stage as shown in Fig. 2, the azimuthal ${\bf E} \times {\bf B_x}$ electron drift channel becomes strongly deformed, and the presence of electron density non-uniformity and $E_{y}$ become much more pronounced. To specify, the electron density is redistributed with its peak in the area where spoke is located. Meanwhile, in the spoke region with higher electron density, the ${\bf E} \times \bf B_x$ drift channel is dragged towards the cathode and the rest of the channel is pushed away from the cathode. The linear analysis using Eq. 2 is not applicable in the nonlinear stage. But, checking the local validity of the instability condition by Eq. 3 can offer insights on the instability saturation. In principle, one has to calculate the sign of $\triangle$ at each data sampling point in the simulation box to precisely distinguish the stability condition locally, particularly in the vicinity of the spoke. But it is challenging to do so due to the noise of $n_e$, whose derivative can produce various glitches. Here, we choose two azimuthal positions where the axial data are sampled and one axial position where the azimuthal data are sampled to evaluate the system stability. Two azimuthal locations are those where the electron density peak and trough are located ($y=2.9 \,{\rm cm}$ and $y=1 \,{\rm cm}$ at $t=t_4$, $y=1 \,{\rm cm}$ and $y=2.8 \,{\rm cm}$ at $t=t_5$ as shown by the dashed lines in Fig. 3); the axial location is at $z=0.65 \,{\rm cm}$ where the electron density is peaked. We should stress that along the axial lines in the two azimuthal positions, $\triangle_{y}$ can be neglected compared to $\triangle_z$, namely $|\triangle_{y}| \ll |\triangle_{z}|$. The axial profiles of $n_e$, $B_x$ and $E_z$ at the two azimuthal locations and at $t=t_4$ and $t=t_5$ are plotted in Fig. 7a-7d. We note the shaded regions in these graphs denoting the spoke front (where plasma is deformed the most) are of interest to check the instability criteria. At the $n_e$ peak position for both moments in the shaded region, $L_n>0$ and $L_B<0$ giving $1/L_n-1/L_B>0$, and $E_z<0$ are detected, namely $\triangle<0$. At the $n_e$ trough for both moments at the shaded region, $E_z<0$, $L_n\approx -0.9 \,{\rm cm}$ and $L_B \approx -1.6 \,{\rm cm}$, meaning $1/L_n-2/L_B>0$ and $\triangle<0$. In terms of the chosen axial position at $z=0.65 \,{\rm cm}$, $\triangle_z<0$ applies along the azimuthal direction. Therefore, we only need to check the sign of $\triangle_{y}$. The azimuthal profiles of $n_e$ and $E_{y}$ are displayed in Fig. 7e and 7f. Like in Fig. 6, $M$ and $N$ are marked to represent the cross points between vertical lines of $n_e$ peak and trough and the horizontal line of $E_{y}=0$. At $t=t_4$ and $t=t_5$, one can tell that $M$ and $N$ overlap two points of the $E_{y}$ curve. At $t=t_4$, on both left side and right side of $M$ and $N$, $\triangle_{y}<0$ applies. At $t=t_5$, on two sides of $N$ and the right side of $M$, $\triangle_{\theta}<0$ is identified. But, on the left side of $M$, the small scale (wavelength $\lambda\approx 3 \,{\rm mm}$) oscillation is seen around $E_{y}=0$ and hence it is hard to tell the sign of $\triangle_{y}$ locally. But on average, $E_{y} \lesssim 0$ on the left of $M$ and $L_n>0$, meaning $\triangle_{y}<0$ in that region. In the animation of ion density in the supplementary material, the oscillation present in the spoke rear can also be clearly identified. It is noteworthy that this small-scale fluctuation does not change the spoke dynamics but coexists with the spoke mode. The oscillation could be due to kinetic instability or ion sound instability found in the magnetron experiments \cite{Tsikata2015}, and its detailed study is out of the scope of this paper and will be studied in the future.

Therefore, in terms of the large-scale spoke mode, the instability condition is not met in the chosen three locations in the nonlinear saturated stage. This leads us to propose that the GDI saturation is caused by the destruction of the instability condition due to the redistribution of the electron density and the deformation of the potential. Consequently, the potential deformation results in the formation of the spoke potential hump, surrounding which the azimuthal $E_{y}$ is present.

To explain the RF-modulated spoke ionization shown in Fig. 1f, the transient characteristics of the plasma quantities are studied. Fig. 8 gives the 2D maps of $\varphi$ (potential), $E_{y}$, $E_z$, $T_{e\perp}$ (electron temperature perpendicular to the magnetic field) and $n_e\nu$ (electron-neutral ionization rate) at four snapshots of one RF period $t/T=0,\pi/2,\pi,3\pi/2$. The snapshots are marked in Fig. 9a. From the maps of $T_{e\perp}$ and $n_e\nu$, the RF-modulated ionization is confirmed with the observation that $T_{e\perp}$ and $n_e\nu$ are peaked at $t/T=3\pi/2$, and then decreases subsequently at $t/T=0, \pi/2, \pi$. Meanwhile, from the maps of $\varphi$, $E_{y}$ and $E_z$, it is interesting to see that the potential hump and the deformed channel hold during the whole RF period, thereby leading to the non-zero $E_{y}$ surrounding the potential hump. This means the electrons in the spoke (potential hump) front are subject to the $\bf E \times \bf B_x$ electron drift along the equipotential lines, as well as the $\nabla B_x$ drift leading to the heating or cooling depending on the sign of $E_{y}$ according to Eq. 1. Fig. 9a further gives the temporal evolution of $E_{y}$, $E_z$ at a fixed point of the spoke front where $E_{y}$ is peaked. The point is marked in the $T_{e\perp}$ plot in Fig. 8. The temporary change of cathode voltage is also presented in Fig. 8a. One important thing as seen in Fig. 8 and Fig. 9a is that the $E_{y}$ amplitude is time changing, which shows maxima at $t/T=3\pi/2$ and minima at $t/T=\pi/2$. According to Eq. 1, the time-changing $E_{y}$ can explain the phenomena of RF-modulated spoke ionization. However, in the cases under study, the electron collision frequency $f_{coll} \approx 5\times 10^6 \,{\rm MHz}$ is less than or comparable to the radio frequency $f=13.56 \,{\rm MHz}$, so the collisional dissipation could not explain two questions: how is the $\nabla B_x$-induced energy gain deposited to the "thermal energy" at $t=3\pi/2$ (increase of $T_{e\perp}$) and then where does the obtained energy go (decrease of $T_{e\perp}$) at $t/T=0, \pi/2,\pi$?    

\begin{figure}
\center
\includegraphics[clip,width=0.9\linewidth]{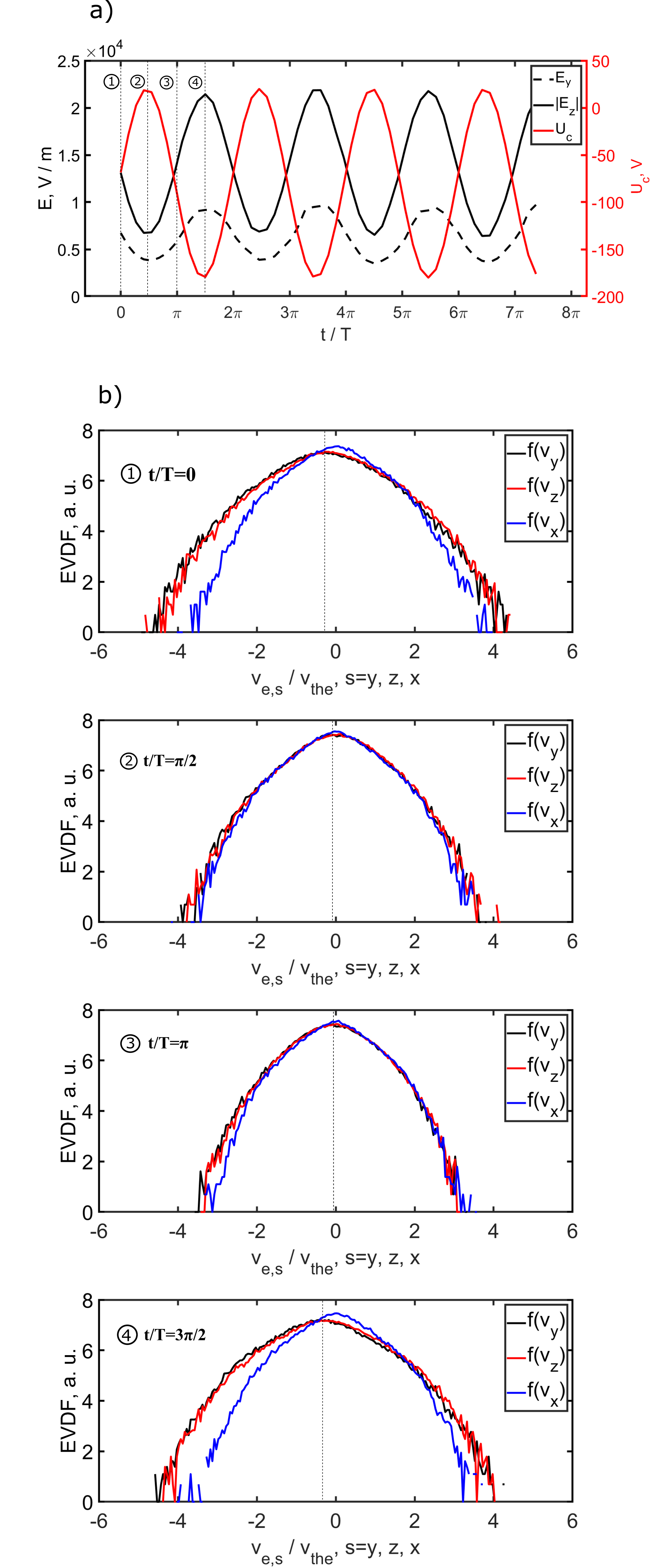}
\caption{a) shows the temporal evolution of $E_{y}$, $|E_z|$, and $U_c$. Note $E_{y}$ and $|E_z|$ are obtained at the point marked in the $T_{e\perp}$ plot shown in Fig. 8 and $U_c$ is the cathode voltage. b) gives EVDFs inside the spoke for three dimensions and at the four snapshots of one RF period. The electrons generating the EVDFs are collected in the spoke position denoted by the box marked in the $T_{e\perp}$ plot shown in Fig. 8. In the figures, $v_{the}$ is a scaled electron thermal velocity $v_{the}=1.0\times 10^6 \,{m/s}$ by $T_e=3 \,{\rm eV}$.}
\end{figure} 

First, we should emphasize that the energy gain expressed by Eq. 1 is to a large extent converted into the rotational energy, rather than the kinetic energy of the directed guiding center's drift motion. Due to the cyclotron motion, the rotational energy is redistributed equally in $y$ and $z$ direction, effectively meaning the broadening of the electron velocity distribution function in the plane perpendicular to the magnetic field. This explains the large value of $T_{e\perp}$ at $t/T=3\pi/2$ in Fig. 8. In addition, the electrons heated at the spoke front are subject to the $\bf E \times \bf B_x$ drift along the equi-potential lines, which can subsequently take away the rotational energy towards the downstream. So this can explain the decrease of $T_{e\perp}$ at $t/T=0, \pi/2,\pi$. To further look into the electron dynamics, the electron velocity distribution functions (EVDF) at each direction ($x$, $y$ and $z$) and at each snapshot are presented in Fig. 9b. We note the particles representing the EVDF are collected from the spoke region denoted by the box shown in the $T_{e\perp}$ plot in Fig. 8. It is seen at $t/T=3\pi/2$, $f(v_{y})$ and $f(v_{z})$ overlaps and are broadened in comparison with $f(v_x)$, and also shifted by the $\bf E_z\times \bf B_x$ drift about $v_E\approx0.3 v_{the}=3\times 10^5 \,{\rm m/s}$. Here, $v_{the}=1\times 10^6 \,{\rm m/s}$ is the electron thermal velocity with $T_e=3 \,{\rm eV}$. So the drift distance of the electrons in one RF period can be estimated to be $L_E=v_E\times T=2.2 \,{\rm cm}$, two times larger than the spoke length $L_s\approx 1cm$. This proves the electrons in the spoke front heated at around $t/T=3\pi/2$ drift away subsequently at $t/T=0,\pi/2,\pi$ leading to the decrease of the local $n_e\nu$ and $T_{e\perp}$. The EVDFs of newly replenished electrons in the spoke position at $t/T=\pi/2, \pi$ show isotropic ones as shown in Fig. 8. In addition, based on Eq. 1 and the parameters at around $t/T=3\pi/2$ shown in Fig. 8, the electron energy gain can be $\varepsilon_{\perp} \approx E_{y}\times v_{\nabla B} \times L_s/v_E \approx 4 \,{\rm eV}$. This also agrees with the $T_{e\perp}$ scenario shown in Fig. 8. Therefore, regarding the RF-modulated spoke ionization, the $\nabla B$ drift induced electron heating explains the enhanced ionization at $t/T=3\pi/2$; the ${\bf E} \times {\bf B}_x$ drift induced electron convection causes the decrease of the ionization rate at $t/T=0,\pi/2,\pi$.

The phenomena of RF-modulated spoke ionization indicates that the amplitude of $E_{y}$ must relate to the cathode voltage. As shown in Fig. 9a, $E_{y}$ is oscillating and synchronized with $E_z$ and the cathode voltage $U_c$. It is easy to understand $|E_y(t)|\propto |E_z(t)|$ considering the channel deformation angle is almost temporally unchanged (see the 2D maps of $\varphi$ and $E_z$ in Fig. 8). We can make a simple estimation to interpret the synchronization between $|E_{y}|$ and $|U_c|$. The potential jump $\triangle \varphi$ across the electron ${\bf E} \times {\bf B}_x$ drift channel can be expressed as $\triangle \varphi (t) = \sqrt{{\overline{E_{y}(t)}^2+\overline{E_z(t)}^2}}d$. Here $d$ is the channel width and assumed to be temporally unchanged. $\overline{E_z}$ and $\overline{E_{y}}$ are spatially averaged along the path perpendicular to the channel. Due to $|E_y(t)|\propto |E_z(t)|$, $|\overline {E_{y}(t)}|\propto \triangle\varphi(t) \approx U_c(t)$ is derived. This suggests that the spoke ionization can be manipulated by the cathode voltage in DC magnetrons and pulsed magnetrons.

\section{Conclusion}
The rotating spoke mode experimentally observed in RF magnetron discharges was successfully reproduced by means of 2D axial-azimuthal fully kinetic PIC/MCC approach. The spoke rotates in the $-{\bf E} \times {\bf B}$ direction with the velocity $5 {\rm km/s}$, consistent with the experimental observation. The underlying physics of prominent spoke features in RF magnetron discharge: the potential hump and the RF-modulated ionization, were elucidated to uncover the driving mechanism behind the spoke formation. The analysis of the computational results, aided by the GDI linear theory, reveals that the cathode sheath $E_z$ triggers the gradient drift instability, which is saturated with the destruction of the GDI condition due to the deformation of the potential and the redistribution of the electron density. As a consequence, the potential hump region forms with the presence of the azimuthal $E_{y}$. We found that in the linear stage, the instability mode wavelength and the corresponding growth rate are in good agreement with the prediction of the GDI linear theory. It is further shown that the saturation level of $E_{y}$ is synchronized with and proportional to the cathode voltage $U_c$, leading to the RF-modulation of the spoke ionization. When $|E_{y}|$ is peaked, the local spoke ionization is enhanced by the $\nabla B$ drift induced electron heating. When $|E_{y}|$ is in the trough, the electron convection due to $\bf E \times \bf B$ drift takes away the heated electrons and causes the decrease of the local ionization rate. The synchronization between $E_{y}$ and $U_c$ suggests that the spoke ionization can be manipulated by the cathode voltage and their quantitative relation deserves detailed study towards the establishment of predictive modeling of magnetron sputtering discharges.

In this paper, we limited ourselves to 2D axial-azimuthal simulations, the self-consistent self-bias DC voltage in the experiments is not reproducible in the chosen geometry. The applied DC voltage in the simulation can introduce the DC current and the electric field, which are absent in the experiments. However, the GDI instability in the simulations was found to be triggered by the cathode sheath electric field, so the DC current induced electric field is expected not to largely change the spoke dynamics. The role of the self bias due to geometrical asymmetry on the spoke formation is reserved for the future study by means of 3D code.


\section*{Acknowledgement}
L Xu gratefully acknowledges the support of the startup funding of Soochow University. S Ganta, I Kaganovich, K Bera and S Rauf were funded by the USA Department of Energy under PPPL-AMAT CRADA agreement, entitled  "Two-Dimensional Modeling of Plasma Processing Reactors". L Xu also would like to express the gratitude to RP Brinkmann for insightful discussions and to M Panjan for communicating the experimental details.

\onecolumn

\appendix\section{Effect of RF electric field on gradient drift instability theory}

To address the effect of the RF electric field on the gradient drift instability (GDI), following to Ref. %
\cite{lakhin2018effects}, also see \cite{smolyakov2016}, we check terms involving the RF oscillation in the equations of the two fluid model used to derive the GDI dispersion relation. We should point out that the purpose is to explain why the RF oscillating electric field is negligible and the RF period averaged electric field can be used as the equilibrium electric field directly to calculate the dispersion relation. Therefore, the full derivation of the GDI dispersion relation will not be presented here, which was detailed in Ref. \cite{lakhin2018effects}. First, we should point out some conditions or approximations in our simulations.

a) The fluctuation frequency of the drift wave is $\omega/2\pi \lesssim 1 \,{\rm MHz}$, much smaller than the RF frequency $\omega_{rf}\approx 13.56 \,{\rm MHz}$. This implies that the electron dynamics and electric field can be split into low frequency component and high frequency component, which will be elaborated below.

b) The electron-neutral ionization collision frequency is about $\nu \approx 0.01 \,{\rm MHz}$, which is much smaller than $\omega$ and $\omega_{rf}$. Hence, the ionization is not considered in the model.

c) The electron cyclotron frequency is on the order of $\omega_{ce} \approx 10 \,{\rm GHz} \gg \omega$. This allows us to expand the electron momentum equation by the smallness $\omega/\omega_{ce}$.

d) The ions do not respond to the RF oscillation, so the RF electric field does not impact ion dynamics and ion equations of the two-fluid model will only be presented in Eq. A1 and A2, and not be discussed subsequently.

The adopted two fluid equation for the description of magnetron plasmas is 

\begin{equation}
     \frac{\partial n_i}{\partial t} +  \triangledown \cdot (n_i \mathbf{v_i}) = 0,
\end{equation}

\begin{equation}
     \frac{\partial \mathbf{v_i}}{\partial t} +  (\mathbf{v_i} \cdot \triangledown)  \mathbf{v_i} = -\frac{e}{m_i}\triangledown \varphi.
\end{equation}

\begin{equation}
     \frac{\partial n_e}{\partial t} +  \triangledown \cdot (n_e \mathbf{v_e}) = 0,
\end{equation}

\begin{equation}
 \begin{aligned}
     \frac{\partial \mathbf{v_e}}{\partial t} + (\mathbf{v_e} \cdot \triangledown)  \mathbf{v_e} = -\frac{e}{m_e}(\mathbf{E}+\mathbf{v_e}\times \mathbf{B_0})  
     -\frac{\triangledown p_e}{n_em_e}-\frac{\triangledown \cdot {\bf \Pi_e}}{n_em_e} 
      \end{aligned}
\end{equation}

\begin{equation}
    \nabla \cdot {\bf E}=-\frac{e}{\varepsilon_0}(n_e-n_i),
\end{equation}

\noindent where $n_i$ is the ion density, $n_e$ the electron density, $\mathbf{v_i}$ ion velocity, $\mathbf{v_e}$ the electron velocity, $m_i$ the ion mass, $m_e$ the electron mass, $\bf E$ the electric field, $\bf B_0$ the external magnetic field, $p_e$ the electron pressure scalar, ${\bf \Pi_e}$ the gyroviscous tensor and $e$ the elementary charge. Eq. A1 and A2 are the mass and momentum conservation equations of ions.  Eq. A3 and A4 are the mass and momentum conservation equations of electrons. The Poisson equation A5 relates the electron density and ion density and closes the model equations' system. 

Let's consider the condition without RF oscillation first. As mentioned, in our simulations, the magnetic field is strong, namely $\omega _{ce} \gg \left( \partial/\partial t\right)$, meaning Eq. A4 can be expanded by the smallness of $\omega/\omega_{ce}$. For the linear dispersion relation derivation, we seek the velocity on the zeroth order and first order by $\omega/\omega_{ce}$

\begin{equation}
\mathbf{v}_{\bot e}=\mathbf{v}_{e}^{(0)}+\mathbf{v}_{e}^{(1)}
\end{equation}%

We note here that the expansion of Eq. A6 is only possible for the velocity
perpendicular to the the magnetic field $\mathbf{v}_{\bot e}$. The lowest leading order electron drift is given by the $\mathbf{E}\times \mathbf{B}$ drift
and diamagnetic drift $\mathbf{v}_{*}$
\begin{equation}
\mathbf{v}_{e}^{(0)}=\mathbf{v}_{E}+\mathbf{v}_{*}=-\frac{1}{B_{0}}\mathbf{b}%
\times {\bf E} -\frac{1}{en_{e0}B_{0}}\mathbf{b}\times
\triangledown p_{e}.  \label{v0}
\end{equation}

\noindent where $\mathbf{b}$ is the unit magnetic field. In the next order, from the momentum balance Eq. A4, we
have the following expression for the electron velocity
\begin{equation}
n_{e}m_{e}\frac{\partial \mathbf{v}_{e}^{(0)}}{\partial t}+n_{e}m_{e}(%
\mathbf{v}_{e}^{(0)}\cdot \triangledown )\mathbf{v}_{e}^{(0)}=-en_e\mathbf{v}%
_{e}^{(1)}\times \mathbf{B}_{0}-\triangledown \cdot \mathbf{\Pi }_{g}.
\end{equation}

Following Eq. 13 of Ref. \cite{lakhin2018effects}, the first order velocity can be expressed as

\begin{eqnarray}
\begin{aligned}
{\bf v_e^{(1)}}=\frac{1}{\omega_{ce}} 
\{ 
\left[\frac{\partial}{\partial t}+({\bf v_E}+{\bf v_{\nabla B}})\cdot \nabla\right] {\bf v}_e^{(0)}+\frac{1}{m_en_e}\nabla \times \left[\frac{p_e}{2\omega_{ce}}(\nabla\cdot {\bf v_e^{(0)}}){\bf b}\right]\\+\frac{1}{m_en_e}\nabla\left[\frac{p_e}{2\omega_{de}}{\bf b}\cdot(\nabla \times {\bf v_e^{(0)}})\right] 
\} \times {\bf b}
\end{aligned}
\end{eqnarray}

Therefore, the objective in this Appendix turns out to be checking whether RF oscillation has impact on $\bf v_e^{(0)}$ and $\bf v_e^{(1)}$. Now let's consider the condition of RF oscillation system. As mentioned, the instability frequency of the drift wave modes of interest is on the magnitude of $\rm MHz$, which is much smaller than the applied radio frequency here $\omega_{rf}=13.56 \,{\rm MHz}$. To show the effects of RF components, we apply the multiple time-scale separation approach to split the electric field and electron quantities into high- and low-frequency components \cite{bellan_2006,Sun2022PRE,Sun2022PRL}:

\begin{equation}
    {\bf E}={\bf E}_l+{\bf E}_h,
\end{equation}

\begin{equation}
    n_e=n_{el}+n_{eh},
\end{equation}

\begin{equation}
    {\bf v}_e={\bf v}_{el}+{\bf v}_{eh},
\end{equation}

\begin{equation}
   p_e=p_{el}+p_{eh},
\end{equation}

\begin{equation}
    {\bf \Pi}_e={\bf \Pi}_{el}+{\bf \Pi}_{eh},
\end{equation}

\noindent where $X_l$ ($X=n_e, {\bf E}, {\bf v_e}, p_e, {\bf \Pi_e}$) is defined as $X_l=X_0+X_1=\overline{X(y,z,t)}=1/T \int_0^T X(y, z,t)dt$ where T is the RF period. $X_0$ and $X_1$ are the equilibrium component and perturbed component of $X$ at the frequency of drift waves, which are quantities both having much lower frequency compared to RF oscillating component $X_h$. Hence, $X_h=X-X_l$. Now we apply RF period averaging to Eqs. A3-A5 and obtain the low-frequency components of electron equations and Poisson equation

\begin{equation}
     \frac{\partial n_{el}}{\partial t} +  \triangledown \cdot (n_{el} \mathbf{v}_{el}+\overline{n_{eh} \mathbf{v}_{eh}}) = 0,
\end{equation}

\begin{equation}
 \begin{aligned}
     \frac{\partial \mathbf{v}_{el}}{\partial t} +  (\mathbf{v}_{el} \cdot \triangledown)  \mathbf{v}_{el} 
     +\overline{(\mathbf{v}_{eh} \cdot \triangledown)  \mathbf{v}_{eh}} 
     = -\frac{e}{m_e}(\mathbf{E}_l+\mathbf{v}_{el}\times \mathbf{B_r})  
     -\frac{\triangledown p_{el}}{n_{e}m_e}-\frac{\triangledown \cdot {\bf \Pi}_{el}}{n_em_e}
      \end{aligned},
\end{equation}

\begin{equation}
    \nabla \cdot {\bf E}_l=-\frac{e}{\varepsilon_0}(n_{el}-n_i),
\end{equation}

Correspondingly, for high frequency component, the electron equations and the Poisson equation write

\begin{equation}
     \frac{\partial n_{eh}}{\partial t} + \triangledown \cdot (n_{eh} \mathbf{v}_{eh}+n_{el} \mathbf{v}_{eh}+n_{eh} \mathbf{v}_{el}-\overline{n_{eh} \mathbf{v}_{eh}}) = 0,
\end{equation}

\begin{equation}
 \begin{aligned}
     \frac{\partial \mathbf{v}_{eh}}{\partial t} + (\mathbf{v}_{eh} \cdot \triangledown)  \mathbf{v}_{eh} 
     -\overline{(\mathbf{v}_{eh} \cdot \triangledown)  \mathbf{v}_{eh}} 
     = -\frac{e}{m_e}(\mathbf{E}_h+\mathbf{v}_{eh}\times \mathbf{B_r})  
     -\frac{\triangledown p_{eh}}{n_em_e}-\frac{\triangledown \cdot {\bf \Pi}_{eh}}{n_e m_e}
      \end{aligned},
\end{equation}

\begin{equation}
    \nabla \cdot {\bf E}_h=-\frac{e}{\varepsilon_0}n_{eh},
\end{equation}

As seen in Eq. A15 and A16, it is apparent that the RF oscillation introduces an additional term $\nabla \cdot (\overline{n_{eh}{\bf v}_{eh}})$ in Eq. A15 and an additional term $\overline{(\mathbf{v}_{eh} \cdot \triangledown)  \mathbf{v}_{eh}}$ in Eq. A16. We estimate the term $\overline{(\mathbf{v}_{eh} \cdot \triangledown)  \mathbf{v}_{eh}}$ first, by using Eq. A19. On the scale of RF frequency, terms $\nabla p_{eh}/n_em_e$, $\nabla \cdot {\bf \Pi_{eh}}/n_em_e$ are negligible compared to other terms, hence Eq. A19 can be approximated to be

\begin{equation}
     \frac{\partial \mathbf{v}_{eh}}{\partial t} + \frac{e}{m_e}\left(\mathbf{E}_{h}+\mathbf{v}_{eh}\times \mathbf{B_0}\right)=0  
\end{equation}

The we get

\begin{equation}
     |v_{eh}|=\left|\frac{ieE_h}{m_e(\frac{\omega_{ce}^2}{\omega_{rf}}-\omega_{rf})}\right|
\end{equation}

This hence gives $\bf A$ defined by 

\begin{equation}
     {\bf A}=\frac{m_e}{e}\overline{(\mathbf{v}_{eh} \cdot \triangledown)  \mathbf{v}_{eh}}=\frac{m_e}{2e}\nabla\left|\overline{v_{eh}^2}\right|=\frac{e\nabla E_{h,max}^2}{4m_e(\frac{\omega_{ce}^2}{\omega_{rf}}-\omega_{rf})^2}
\end{equation}

\noindent where $E_{h,max}$ is the electric field amplitude of the RF oscillation. Let's estimate the value of A. In our cases, $E_{h,max} \approx E_l \approx 10^4 \,{\rm V/m}$ (see Fig. 9a). If we use the $n_e$ gradient length $L_n \approx 1 \,{\rm cm}$ to approximate $\nabla E_{h, max}^2$, $\nabla E_{h, max}^2 \approx 10^{12} \,{\rm V^2/m^3}$, giving $A\approx 10^{-2} \,{\rm V/m}$. So the electric field $E_l\approx 10^4 \,{\rm V/m} \gg A$ means the effect of RF electric field is negligible on the zeroth order electron velocity. Further, according  Eq. A9 and A16, considering the effect of RF oscillation, the electron velocity on the first order is expressed as

\begin{equation}
\begin{aligned}
{\bf v_e^{(1)}}=\frac{1}{\omega_{ce}}\{\left[\frac{\partial}{\partial t}+({\bf v_E}+{\bf v_D})\cdot \nabla\right] {\bf v_e^{(0)}}+\frac{1}{m_en_e}\nabla \times \left[\frac{p_e}{2\omega_{ce}}(\nabla\cdot {\bf v_e^{(0)}}){\bf b}\right]+\\ \frac{1}{m_en_e}\nabla\left[\frac{p_e}{2\omega_{ce}}{\bf b}\cdot(\nabla \times {\bf v_e^{(0)}})\right]+\frac{e{\bf A}}{m_e}\}\times {\bf b}
\end{aligned}
\end{equation}

The last term on the right hand of Eq. A24 represents the effect of RF electric field. We can insert $\bf v_e^{(1)}$ to electron continuum equation Eq. A15 (for the purpose of linearization), we can get an additional term regarding the RF oscillation

\begin{equation}
\frac{1}{\omega_{ce}}{\frac{en_e}{m_e}\nabla \cdot ({\bf A}\times {\bf b}})=\frac{1}{\omega_{ce}}{\frac{en_e}{m_e}}{\bf b}\cdot (\nabla \times {\bf A}) = 0
\end{equation}

This clearly suggests that the RF oscillation present in the first order velocity $\bf v_e^{(1)}$ can also be negligible for the linearization at the next step. Therefore, in terms of the electron momentum equation, the high frequency (RF) component can be excluded in the derivation of the GDI dispersion relation. Finally, we check the electron continuum equation as shown in Eq. A15, where the term of $\nabla \cdot (\overline{n_{eh}{{\bf v}_{eh}}})$ may introduce the RF ocsillation effect. 

Combining Eq. A20 and Eq. A22, we can obtain

\begin{equation}
\nabla \cdot (\overline{n_{eh} {{\bf v}_{eh}}})=-\frac{i\epsilon_0\nabla^2 \overline{E_{h}^2}}{2m_e(\frac{\omega_{ce}^2}{\omega_{rf}}-\omega_{rf})}\approx -\frac{i\epsilon_0\omega_{rf}\nabla^2 \overline{E_{h}^2}}{2m_e\omega_{ce}^2}
\end{equation}

For the linearization, then substitute Eq. A7, A9 and A26 into Eq. A15, and consider the local perturbed plasma quantities $\widetilde{X_l}\sim exp[-i(\omega t-k_{y}y-k_{z}z)]$, following to Eq. 18 of Ref. \cite{lakhin2018effects}, we can linearize Eq. A15 in the Fourier space to be 

\begin{equation}
\begin{aligned}
\left[(\omega-\omega_E-\omega_D)+k_{\perp}^2\rho_e^2(\omega-\omega_E-2\omega_D)\right]\widetilde{n_e}+\frac{\epsilon_0\omega_{rf} \overline{E_{h}^2}}{2m_e\omega_{ce}^2}k_{\perp}^2\\
=\left[\omega_*-\omega_D+k_{\perp}^2\rho_e^2(\omega-\omega_E-2\omega_D)\right]\frac{en_e}{T_e}\widetilde{\varphi}
\end{aligned}
\end{equation}

\noindent where $\omega_E=k_{y}v_E$, $\omega_*=k_{y}v_{*}$, $\omega_D=k_{y}v_D$, $v_{D}=2T_e\nabla ln B_0/eB_0$ and $k_{\perp}^2=k_y^2+k_z^2$. In our simulations, $k_z \ll k_y$, $k_{\perp}\rho_e \approx 0.02$, the electron density oscillation $\widetilde{n_e}\approx 10^{15} \,{\rm m^{-3}}$ (see Fig. 6 in the main text), which enable us to estimate

\begin{equation}
\frac{\epsilon_0\omega_{rf} \overline {E_{h}^2}}{2m_e\omega_{ce}^2}k_{\perp}^2 \approx 10^{16} \,{\rm rad/m^{3}s}
\end{equation}

\begin{equation}
\omega \widetilde{n_e}\approx 10^{21} \,{\rm rad/m^{3}s}
\end{equation}

\begin{equation}
k_{\perp}^2\rho_e^2\omega \widetilde{n_e}\approx 10^{18}  \,{\rm rad/m^{3}s}
\end{equation}

This clearly shows that the term involving the RF electric field is less than $\omega \widetilde {n_e}$ and $k^2\rho_e^2 \omega \widetilde {n_e}$ by at least two orders of magnitude, meaning the RF electric field also has less effect on the electron continuum equation. It is hence summarized that the RF electric field component $E_h$ can be neglected to derive the GDI dispersion relation, in which the RF period averaged value of electric field can be used as the equilibrium electric field ($E_0$ in Eq. 2 of the main text). 

\section*{References}
\bibliography{main.bbl}
\end{document}